# ELECTRODYNAMICAL FORBIDDANCE OF A STRONG QUADRUPOLE LIGHT-MOLECULE INTERACTION IN FULLERENE $C_{60}$ AND ITS WITHDRAWAL IN FULLERENE $C_{70}$

**Vladimir P. Chelibanov[1] Aleksey M. Polubotko[2*]**

[1]State University of Information Technologies, Mechanics and Optics, Kronverkskii 49, 197101 Saint Petersburg, RUSSIA  E-mail: Chelibanov@gmail.com

[2] A.F. Ioffe Physico-Technical Institute, Politechnicheskaya 26, 194021 Saint Petersburg, RUSSIA E-mail: alex.marina@mail.ioffe.ru

## Abstract

It is demonstrated that in fullerene $C_{70}$, which can be considered as a deformed fullerene $C_{60}$ in some sense, there is a withdrawal of an electrodynamical forbiddance of a strong quadrupole light-molecule interaction, which is realized in the fullerene $C_{60}$. This situation occurs because of the reduction of symmetry of $C_{70}$ from the icosahedral symmetry group $Y_h$ to the group $D_{5h}$. The withdrawal results in appearance of the lines in the SERS spectra of $C_{70}$, which are forbidden in usual Raman scattering and are allowed in infrared absorption, while such lines are forbidden in the SERS spectrum of the fullerene $C_{60}$ due to the electrodynamical forbiddance. The measured SERS spectra of $C_{70}$ demonstrates existence of such lines that strongly confirms our ideas about the dipole –quadrupole SERS mechanism.

*corresponding author

## Introduction

Investigation of fullerenes by Surface Enhanced Raman scattering is of a great interest since it can be a promising technic for their study and can give us a new information about their structure, optical and some other important properties. As it was demonstrated in [1] the SERS mechanism is well viewed and confirmed in fullerene $C_{60}$ (Fig. 1a). Its main feature in this molecule is an electrodynamical forbiddance of a strong quadrupole light-molecule interaction, which well manifests in its SERS spectrum [1]. The electrodynamical forbiddance means that the quantum mechanical operator of the strong quadrupole light-molecule interaction is equal to zero because of belonging of $C_{60}$ to the icosahedral symmetry group $Y_h$ and due to the electrodynamical law $div\mathbf{E} = 0$ [1]. The main peculiarity of the SERS spectrum of $C_{60}$ is the absence of the lines, caused by vibrations, which refer to the three dimensional irreducible representation $T_{1u}$ (Fig. 2), which describes transformational properties of the dipole moment components $d_{e,x}, d_{e,y}$ и $d_{e,z}$. These lines are forbidden in a usual Raman scattering and are allowed in infrared absorption in molecules with sufficiently high symmetry. However they become allowed in the SERS spectra of such molecules. In $C_{60}$ these lines also become forbidden, but due to the electrodynamical forbiddance [1].

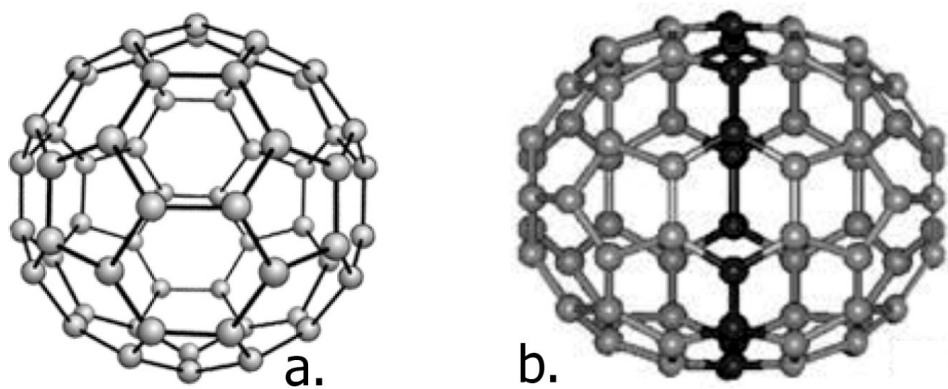

**Figure 1.** a. Fullerene $C_{60}$, b. Fullerene $C_{70}$

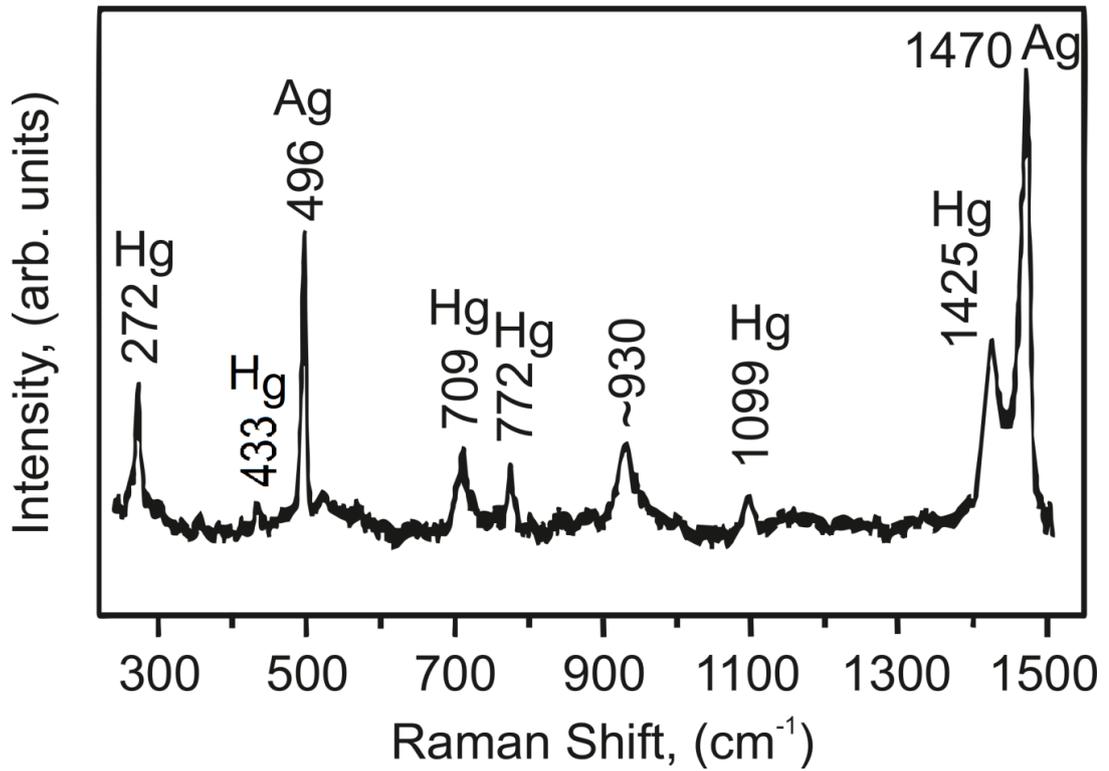

**Figure 2.** The SERS spectrum of the fullerene $C_{60}$. One can see that there are only the lines, caused by irreducible representations $A_g$ and $H_g$. The lines with the irreducible representation $T_{1u}$ are absent in the spectrum because of the electrodynamical forbiddance.

Here we want to pay attention of the readers that in fullerene $C_{70}$, (Fig. 1b), which can be considered as a deformed fullerene $C_{60}$ in some sense, the electrodynamical forbiddance disappears because of the symmetry lowering and the change of its group from $Y_h$ to $D_{5h}$. This manifests in its SERS spectra by appearance of the lines, which are forbidden in usual Raman scattering and are allowed in infrared absorption.

## Peculiarities of the light-molecule interaction Hamiltonian in fullerenes $C_{60}$ and $C_{70}$

As it was demonstrated in our works (see [2] for example), the enhancement in SERS is associated with so-called strong dipole and quadrupole light-molecule interactions, which arise in surface electromagnetic fields existing near rough metal surfaces and in general, near rough surfaces of semiconductors and dielectrics too [3]. In metals the enhancement of the dipole interaction occurs because of the enhancement of the $E_z$ component of the electric field, which is perpendicular to the surface, while the strong quadrupole interaction arises due to the strong increase of the derivatives of the electric field of the $\frac{\partial E_i}{\partial x_i}$ type and due to some features of matrix elements of the quadrupole moments of $Q_{e,ii}$ type, which are of a constant sign. The last feature is associated with a pure quantum mechanical properties of molecules and a matter in general. The moments $Q_{e,ii}$ are responsible for the strong quadrupole light-molecule interaction, which arises in this system and they are named as main quadrupole moments $Q_{main}$. The general form of the light-molecule interaction Hamiltonian can be written as

$$\hat{H}_{e-r} = |\mathbf{E}| \frac{(\mathbf{e}^*\mathbf{f}_e^*)e^{i\omega t} + (\mathbf{e}\mathbf{f}_e)e^{-i\omega t}}{2} \quad , \tag{1}$$

where

$$f_{e,i} = d_{e,i} + \frac{1}{2E_i} \sum_k \frac{\partial E_i}{\partial x_k} Q_{e,ik} \tag{2}$$

is an $i$ component of the generalised coefficient of light-molecule (light-electron) interaction,

$$d_{e,i} = \sum_\alpha ex_{\alpha,i}$$

$$Q_{e,i,k} = \sum_\alpha ex_{\alpha i} x_{\alpha k} \tag{3}$$

are the components of the dipole and quadrupole moments of electrons of the molecule. Here $(i,k) = (x, y, z)$.

In symmetrical molecules of various groups of symmetry, in general case, the moments $Q_{e,ii}$ do not transform after irreducible representations. Therefore it is convenient to transfer from the moments $Q_{e,ii}$ to their linear combinations $Q_{e,1}, Q_{e,2}$ and $Q_{e,3}$, transforming after irreducible representations, which are of the following form

$$Q_{e,1} = b_{11}Q_{e,xx} + b_{12}Q_{e,yy} + b_{13}Q_{e,zz} \tag{4}$$

$$Q_{e,2} = b_{21}Q_{e,xx} + b_{22}Q_{e,yy} + b_{23}Q_{e,zz} \tag{5}$$

$$Q_{e,3} = b_{31}Q_{e,xx} + b_{32}Q_{e,yy} + b_{33}Q_{e,zz} \tag{6}$$

The specific form of the coefficients $b_{ik}$ depends on the symmetry group of the molecule.

The moments $Q_{e,xx}, Q_{e,yy}$ and $Q_{e,zz}$ can be expressed via the moments $Q_{e,1}, Q_{e,2}$ and $Q_{e,3}$ by the following manner

$$Q_{e,xx} = a_{11}Q_{e,1} + a_{12}Q_{e,2} + a_{13}Q_{e,3} \tag{7}$$

$$Q_{e,yy} = a_{21}Q_{e,1} + a_{22}Q_{e,2} + a_{23}Q_{e,3} \tag{8}$$

$$Q_{e,zz} = a_{31}Q_{e,1} + a_{32}Q_{e,2} + a_{33}Q_{e,3} \tag{9}$$

Here the coefficients $a_{ik}$ depend on the specific symmetry group also. There are linear combinations $Q_{e,1}, Q_{e,2}$ and $Q_{e,3}$ with a constant sign, which transform after the unit irreducible representation of the symmetry group and are essential for the scattering, and the combinations with a changeable sign, transforming after other irreducible representations, which are nonessential for the strong scattering. The former we shall name as main quadrupole moments $Q_{main}$, while the latter, the minor ones $Q_{\min or}$, Appearance of the main quadrupole moments, which have another transformational properties than the dipole moments and are essential for the scattering results in appearance of forbidden lines [2].

In symmetrical molecules with the symmetry groups $T, T_d, T_h, O, O_h$ and also in the fullerene $C_{60}$, which belongs to the icosahedral group $Y_h$, the linear combinations $Q_{e,1}, Q_{e,2}$ and $Q_{e,3}$ have the following form

$$Q_{e,1} = \frac{1}{3}(Q_{e,xx} + Q_{e,yy} + Q_{e,zz}) \quad , \tag{10}$$

$$Q_{e,2} = \frac{1}{2}(Q_{e,xx} - Q_{e,yy}) \quad , \tag{11}$$

$$Q_{e,3} = \frac{1}{4}(Q_{e,xx} + Q_{e,yy} - 2Q_{e,zz}) \quad . \tag{13}$$

The main moment in this case is $Q_{e,1}$, which transforms after the unit irreducible representation, while the moments $Q_{e,2}$ and $Q_{e,3}$ transform after other irreducible representations and are the minor ones. The corresponding expressions for $Q_{e,xx}, Q_{e,yy}$ and $Q_{e,zz}$ are the following

$$Q_{e,xx} = Q_{e,1} + \frac{2}{3}Q_{e,3} + Q_{e,2} \quad , \tag{14}$$

$$Q_{e,yy} = Q_{e,1} + \frac{2}{3} Q_{e,3} - Q_{e,2} , \tag{15}$$

$$Q_{e,zz} = Q_{e,1} + \frac{2}{3} Q_{e,3} - 2Q_{e,2} . \tag{16}$$

The value $|\mathbf{E}|(\mathbf{ef}_e)$ in the expression (1) for the light-molecule interaction Hamiltonian for $C_{60}$ can be represented in the form

$$|\mathbf{E}|(\mathbf{ef}_e) = (\mathbf{Ed_e}) + \frac{1}{2} div\mathbf{E} \times \left( Q_{e,1} + \frac{2}{3} Q_{e,3} \right) + \frac{1}{2} \left( \frac{\partial E_x}{\partial x} - \frac{\partial E_y}{\partial y} - 2 \frac{\partial E_z}{\partial z} \right) Q_{e,2}$$
$$+ \frac{1}{2} \sum_{\substack{i,k \\ i \neq k}} \frac{\partial E_i}{\partial x_k} Q_{e,ik}$$
$$\tag{17}$$

One can see that the term with the main quadrupole moment $Q_{e,1}$ is equal to zero due to the factor $div\mathbf{E} = 0$, which is an electrodynamical law. This is the electrodynamical forbiddance of the strong quadrupole light-molecule interaction, which arises due to some features of the strong quadrupole light-molecule interaction and due to belonging of $C_{60}$ to the icosahedral symmetry group. Since all other terms of the quadrupole interaction contain minor quadrupole moments, then the indicated optical processes are determined only by the dipole interaction and their spectra do not contain forbidden lines. For the fullerene $C_{70}$ the expressions for the moments $Q'_{e,1}, Q'_{e,2}$ and $Q'_{e,3}$ have the form

$$Q'_{e,1} = \frac{1}{2}(Q_{e,xx} + Q_{e,yy}) \tag{18}$$

$$Q'_{e,2} = \frac{1}{2}(Q_{e,xx} - Q_{e,yy}) \tag{19}$$

$$Q'_{e,3} = Q_{e,zz} \tag{20}$$

Here we apply the hatched designations for the moments in order to differ the moments for the fullerenes of various types. The irreducible representations, characters and corresponding combinations of the moments for the symmetry group $D_{5h}$ which describes the symmetry properties of the fullerene $C_{70}$ are presented in Appendix.

Now we shall consider the light-molecule interaction Hamiltonian (1) for the fullerene $C_{70}$. The corresponding value $|\mathbf{E}|(\mathbf{ef}_e)$ in (1) in this case has the form

$$|\mathbf{E}|(\mathbf{ef}_e) = (\mathbf{Ed_e}) + \frac{1}{2}\left(\frac{\partial E_x}{\partial x} - \frac{\partial E_y}{\partial y}\right)Q'_{e,2} + \frac{1}{2}\frac{\partial E_z}{\partial z}Q'_{e,3} - \frac{1}{2}\frac{\partial E_z}{\partial z}Q'_{e,1} +$$

$$+ \frac{1}{2}\sum_{\substack{i,k \\ i \neq k}} \frac{\partial E_i}{\partial x_k} Q_{e,ik} \quad . \quad (21)$$

The moments $Q'_{e,2}$ and $Q_{e,ik}$ in the right part of (21) are the minor ones and should not influence on the enhancement. The third and the forth terms contain the main moments, which must cause the enhancement due to the strong quadrupole interaction. In case we shall transfer to the icosahedral symmetry group $Y_h$ they will not cause any enhancement because the moments $Q_{e,xx}, Q_{e,yy}$ and $Q_{e,zz}$ in this case transform one through another and are equivalent. Moreover the value $(Q'_{e,1} - Q'_{e,3})$ in fact coincides with the minor moment $Q_{e,3}$, which is determined to the icosahedral symmetry group $Y_h$ (13) with the accuracy of the factor $\frac{1}{2}$. Therefore this transformation is the one, when all the quadrupole moments become the minor ones.

## Peculiarities of the SERS spectrum of the fullerene $C_{70}$ and withdrawal of the electrodynamical forbiddance

In accordance with the dipole-quadrupole theory [2], the SERS cross-section for the $s$ vibrational mode is determined by the sum of scattering contributions, which correspond to the scattering via various dipole and quadrupole moments $f_1$ and $f_2$ (Fig. 3)

$$d\sigma_{s,surf} = \frac{\omega_{inc}\omega_{scat}^3}{16\hbar^2\varepsilon_0^2\pi^2 c^4} \frac{|\mathbf{E}_{inc}|^2_{surf}}{|\mathbf{E}_{inc}|^2_{vol}} \frac{|\mathbf{E}_{scat}|^2_{surf}}{|\mathbf{E}_{scat}|^2_{vol}} \times$$

$$\times \sum_p \binom{(V_{(s,p)}+1)/2}{V_{(s,p)}/2} \left|T_{d-d} + T_{d-Q} + T_{Q-d} + T_{Q-Q}\right|^2_{surf} dO \quad . \quad (22)$$

Here $\mathbf{E}_{inc}$ and $\mathbf{E}_{scat}$ are the incident and the scattering electric fields, the signs *surf* and *vol* mean that we consider the fields on the surface and in the volume respectively, $\omega_{inc}$ and $\omega_{scat}$ are the frequencies of these fields, $V_{(s,p)}$ - is the vibrational quantum number of the degenerate vibrational mode $(s,p)$, where $s$ numerates the group of degenerated states, $p$ numerates the states inside the group. $T$ - designates the sum of the contributions of the dipole-dipole, dipole-quadrupole, quadrupole-dipole and quadrupole-quadrupole scatterings, $dO$ is the element of a solid angle. Other designations are conventional. Here we do not write out the expressions for the contributions. The reader can find them in the monograph [2]. Every contribution obeys to selection rules

$$\Gamma_{(s,p)} \in \Gamma_{f_1} \times \Gamma_{f_2} . \quad (23)$$

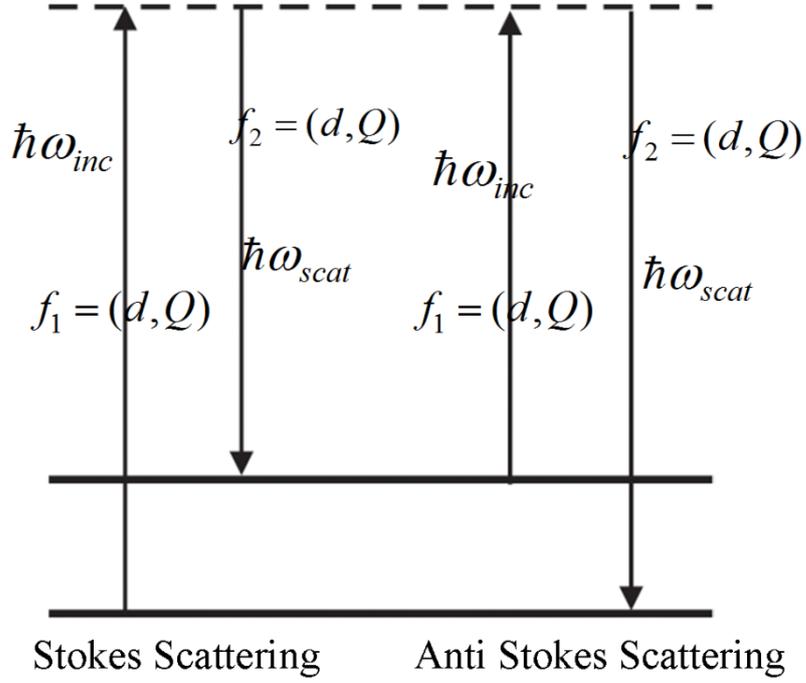

Figure 3. The SERS scattering diagram. The scattering occur via various combinations of the dipole and quadrupole moments $d$ and $Q$.

Here the sign $\Gamma$ designates the irreducible representation, which describe transformational properties of the vibrational mode $(s, p)$ and of the dipole and quadrupole moments $f_1$ and $f_2$. Further we shall designate every contribution, which occur due to the scattering via the dipole and quadrupole moments simply as $(f_1 - f_2)$. It is necessary to note that the relative value of the contributions depends on experimental conditions, or on the surface roughness degree. For the case of a very strong roughness of the substrate, the quadrupole interaction can be significantly more than the dipole ones [2] and the most enhancement experience the contributions of the $(Q_{main} - Q_{main})$ type. In accordance with the selection rules (23), these contributions determine the lines, which are caused by the unit irreducible representation. The contributions of the $(Q_{main} - d_i)$ and $(d_i - Q_{main})$ types are also strongly enhanced, but with a lesser degree than the indicated previous ones. These contributions are caused by the vibrations, which transform such as the dipole moments $d_i$ and determine appearance of the lines, which are forbidden in usual Raman scattering in molecules with sufficiently high symmetry. The contributions of the $(d_i - d_i)$ and $(d_i - d_k)$ types $(i \neq k)$, can be strongly enhanced also, but with a lesser degree than the previous two types. Moreover, when the quadrupole interaction with the main quadrupole moments is not very strong that can be in case not very strong roughness degree, the contribution of the $(d_i - d_i)$ type can define the intensities of some lines, caused by the vibrations transforming after the unit irreducible representation. However, as it was indicated above, in general case, the relative value of the contributions depends on the experimental conditions. Specifically, investigation of the SERS spectra of $C_{70}$ were performed in [4]. In Table 1 one can find the list of the wavenumbers of the SERS spectrum of $C_{70}$ and corresponding irreducible representations. As it follows from these results, the most enhanced are the lines, caused by the vibrations with irreducible representations $E_2'$ and $E_1''$ (Fig. 4), which describe transformational properties of the moments of the $xy, xz$ and $yz$ type and therefore these lines are determined mostly by the enhanced dipole interaction and by the enhanced electric field rather than by the quadrupole interaction. The lines, caused by vibrations with the unit irreducible representation $A_1^{'}$ are enhanced significantly weaker. This fact also indicates that the dipole interaction in this specific case is significantly stronger than the quadrupole one. The strong enhancement of the lines with the irreducible representations $E_2^{'}$ and $E_1^{''}$, is caused by the fact that all the three components of the electric field in the coordinate system, associated with $C_{70}$ can be enhanced. This situation occurs because $C_{70}$ apparently can adsorb on the metal substrate, having arbitrary orientation with respect to the surface and with respect to the enhanced component of the electric field $E_z$, which is perpendicular to the surface. The $E_z$ component is projected on all three axes of the coordinate system, associated with the molecule and there are all three enhanced components of the electric field $E_x^{'}, E_y^{'}$ and $E_z^{'}$. Here we introduce the hatched designations of the electric field components in the coordinate system, associated with the $C_{70}$ molecule that differ them from the components of the field, in the coordinate system, associated with the surface. It is necessary to note that there are two lines with a weak intensity, which refer to vibrations with the irreducible representations $A_2^{''}$ and $E_1^{'}$ and with the wavenumbers at 357 and 1031 $cm^{-1}$, which are forbidden in usual Raman scattering, This result demonstrates appearance of the strong quadrupole light-molecule interaction in this system and withdrawal of the electrodynamical forbiddance. The small intensities of these lines indicate that the quadrupole interaction is sufficiently weak in this system. This fact may be associated with a weak symmetry distortion with respect to the icosahedral group $Y_h$. In addition in accordance with experimental conditions in [4], the metal substrate is the one of a mirror type with the roughness of a characteristic size, which is equal approximately to 100 nm. Apparently the roughness was "sufficiently weak" for these experimental conditions that results not in a strong enhancement of the electric field and its derivatives, when the quadrupole interaction was less than the dipole one.

Table 1. Assignment of the SERS lines of the fullerene $C_{70}$ to irreducible representations of the $D_{5h}$ symmetry group. IR active – are the lines, which are active in infrared absorption.

| Wavenumber ($cm^{-1}$) | Irreducible representations of the $D_{5h}$ symmetry group | Wavenumber ($cm^{-1}$) | Irreducible representations of the $D_{5h}$ symmetry group |
|---|---|---|---|
| 221 | $A_1'$ | 943 | $E_1''$ |
| 254 | $E_2'$ | 990 | $E_2'$ |
| 357 | $A_2''$ IR active | 1031 | $E_1'$ IR active |
| 392 | $E_2'$ | 1056 | $E_2'$ |
| 410 | $E_2'$ | 1180 | $E_1'', E_2'$ |
| 426 | $E_1''$ | 1212 | $E_1'', E_2'$ |
| 450 | $A_1'$ | 1224 | $E_1''$ |
| 500 | $E_2'$ | 1250 | $E_1''$ |
| 533 | $A_1'$ | 1281 | $A_1'$ |
| 564 | $E_1''$ | 1301 | $E_2'$ |
| 574 | $E_1''$ | 1321 | $A_1'$ |
| 610 | $E_2'$ | 1337 | $E_1''$ |
| 654 | $E_1''$ | 1374 | $E_2'$ |
| 697 | $E_2'$ | 1408 | $E_1''$ |
| 718 | $E_1''$ | 1441 | $E_2'$ |

| 734 | $A_1'$ | 1466 | $A_1'$ |
| --- | --- | --- | --- |
| 787 | $E_2'$ | 1509 | $E_1''$ |
| 825 | $A_1'$ | 1563 | $E_2'$ |
| 861 | $E_2'$ | 1583 | $A_1', E_1''$ |
| 893 | $E_1''$ | | |

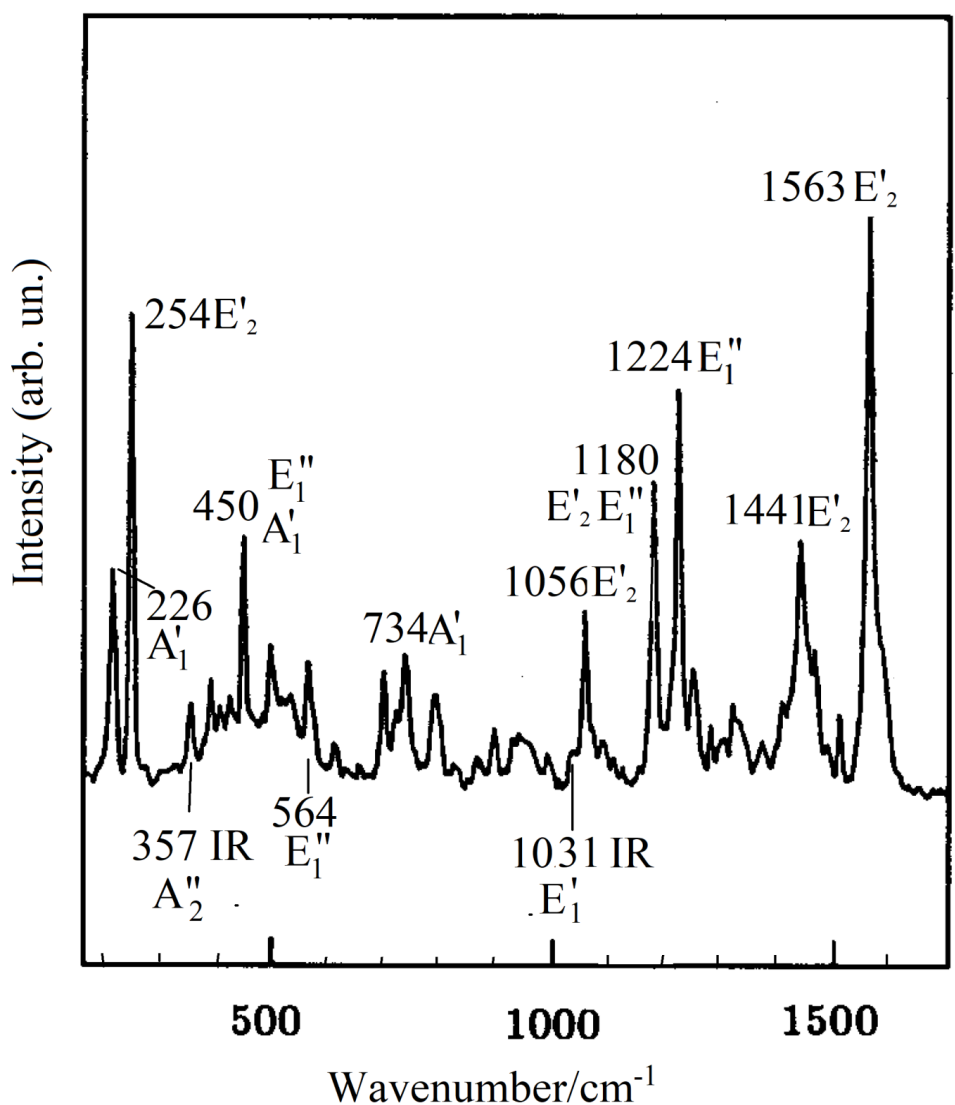

**Figure 4.** The SERS spectrum of the fullerene $C_{70}$. The most enhanced lines are those, caused by the vibrations with the irreducible representations $E_2'$ and $E_1''$, which describe transformational properties of the moments of the $xy, xz$ and $yz$ types.

## Conclusion

Thus because the fullerene $C_{70}$ belongs to the $D_{5h}$ symmetry group and can be considered in some sense as a deformed fullerene $C_{60}$, which is stretched along the $z$ axis, the electrodynamical forbiddance withdraws and the forbidden lines appear in its SERS spectrum. One should note that $C_{70}$ certainly differs from $C_{60}$ by the number of atoms, since it can be considered as $C_{60}$ with some ring, consisting from the rings of carbon atoms, which is introduced in the equator. Therefore its wavenumbers of the vibrations are shifted with respect to the wavenumbers of $C_{60}$. The quadrupole interaction in this specific experiment apparently is sufficiently weak and the strong dipole interaction plays a main role that results in the most enhancement of the lines, caused by the vibrations with the irreducible representations $E_2^{'}$ and $E_1^{''}$. These representations describe transformational properties of the $xy, xz$ and $yz$ moments and this result indicates that the scattering is mostly of the dipole type and is defined mostly by the enhanced electric field. The sufficiently weak quadrupole interaction apparently is associated with the specific experimental conditions, when the roughness degree of the substrate is not strong and also with the fact that the "relative deformation" of $C_{70}$ with respect to $C_{60}$ is sufficiently small. The result that the forbidden lines in $C_{70}$ have sufficiently small intensity corroborates this point of view.

## Appendix

Table 2. Irreducible representations of the $D_{5h}$ group. In the last column of Table 2, one can find linear combinations of the dipole and quadrupole moments which transform after corresponding irreducible representations. The main quadrupole moments in this group are $Q_{e,1}^{'} = Q_{e,xx} + Q_{e,yy}, Q_{e,3}^{'} = Q_{e,zz}$.

| $D_{5h}$ | $C_1$ | $2C_5$ | $2C_5^2$ | $5C_2$ | $\sigma_h$ | $2S_5$ | $2S_5^3$ | $5\sigma_v$ | |
|---|---|---|---|---|---|---|---|---|---|
| $A_1^{'}$ | 1 | 1 | 1 | 1 | 1 | 1 | 1 | 1 | $(Q_{e,xx} + Q_{e,yy}), Q_{e,zz}$ |
| $A_1^{''}$ | 1 | 1 | 1 | 1 | -1 | -1 | -1 | -1 | |
| $A_2^{'}$ | 1 | 1 | 1 | -1 | 1 | 1 | 1 | -1 | |
| $A_2^{''}$ | 1 | 1 | 1 | -1 | -1 | -1 | -1 | 1 | $d_{e,z}$ |
| $E_1^{'}$ | 2 | $2\cos 72^0$ | $2\cos 144^0$ | 0 | 2 | $2\cos 72^0$ | $2\cos 144^0$ | 0 | $(d_{e,x}, d_{e,y})$ |
| $E_1^{''}$ | 2 | $2\cos 72^0$ | $2\cos 144^0$ | 0 | -2 | $-2\cos 72^0$ | $-2\cos 144^0$ | 0 | $(Q_{e,xz}, Q_{e,yz})$ |
| $E_2^{'}$ | 2 | $2\cos 144^0$ | $2\cos 72^0$ | 0 | 2 | $2\cos 144^0$ | $2\cos 72^0$ | 0 | $(Q_{e,xx} - Q_{e,yy}, Q_{e,xy})$ |
| $E_2^{''}$ | 2 | $2\cos 144^0$ | $2\cos 72^0$ | 0 | -2 | $-2\cos 144^0$ | $-2\cos 72^0$ | 0 | |